\documentclass[twocolumns]{aa}%
\usepackage{graphicx}
\usepackage[round]{natbib}
\usepackage{fancyhdr}

\begin{document}
\title{X-ray impact on the protoplanetary disks around T Tauri stars}
\subtitle{}
\author{G.~Aresu\inst{1}
  \and I.~Kamp\inst{1}
  \and R.~Meijerink\inst{2}
  \and P.~Woitke\inst{3,4,5}
  \and W.-F.~Thi\inst{5,6}
  \and M.~Spaans\inst{1}}
\institute{Kapteyn Astronomical Institute, Postbus 800, 9700 AV Groningen, The Netherlands
  \and Leiden Observatory, Leiden University, P.O. Box 9513, NL-2300 RA, Leiden, The Netherlands
  \and UK Astronomy Technology Center, Royal Observatory, Edinburgh, Blackford Hill, Edinburgh EH9 3HJ, UK
  \and Institute for Astronomy, University of Edinburgh, Royal Observatory,  Blackford Hill, Edinburgh EH9 3HJ, UK
  \and Laboratoire d'Astrophisique de Grenoble, CNRS/Universit{\'e} Joseph~Fourier (UMR5571) BP 53, F-38041 Grenoble cedex 9, France
  \and SUPA, Institute for Astronomy, University of Edinburgh, Royal
Observatory Edinburgh, UK
 }
\date{Received 2010 / Accepted 2010}
\abstract {T Tauri stars have X-ray luminosities ranging from $L_{\rm X} = 10^{28}-10^{32}\,\mathrm{erg\,s^{-1}}$. These luminosities are similar to  UV luminosities ($L_{\rm UV} \sim 10^{30}-10^{31} \rm erg\,s^{-1}$) and therefore X-rays are expected to affect the physics and chemistry of the upper layers of their surrounding protoplanetary disks.} {The effects and importance of X-rays on the chemical and hydrostatic structure of protoplanetary disks are investigated, species tracing X-ray irradiation (for $L_{\rm X} \geq 10^{29}\mathrm{erg \,s^{-1}}$) are identified and predictions for [O\,{\sc i}], [C\,{\sc ii}] and [N\,{\sc ii}] fine structure line fluxes are provided.} {We have implemented X-ray physics and chemistry into the chemo-physical disk code ProDiMo. We include Coulomb heating and $\mathrm{H_2}$ ionization as heating processes and primary and secondary ionization due to X-rays in the chemistry.} {X-rays heat up the gas causing it to expand in the optically thin surface layers. Neutral molecular species are not much affected in their abundance and spatial distribution, but charged species such as $\mathrm{N^+}$, $\mathrm{OH^+}$, $\mathrm{H_2O^+}$ and $\mathrm{H_3O^+}$ show enhanced abundances in the disk surface.} {Coulomb heating by X-rays changes the vertical structure of the disk, yielding temperatures of $\sim$ 8\,000 K out to distances of 50 AU. The chemical structure is altered by the high electron abundance in the gas in the disk surface, causing an efficient ion-molecule chemistry. 
The products of this, $\rm OH^+$, $\rm H_2O^+$ and $\rm H_3O^+$, are of great interest for observations of low-mass young stellar objects with the Herschel Space Observatory. [O\,{\sc i}] (at 63 and 145 $\rm \mu m$) and [C\,{\sc ii}] (at $\rm 158\,\mu m$) fine structure emission are only affected for $L_{\rm X} > 10^{30}$ erg s$^{-1}$.}

\maketitle
\section{Introduction}
Young stellar objects emit X-ray radiation (\citealp{Koy94}), which is associated with magnetic coronal processes (\citealp{Fla05,Wol05,Ima01}), jets (\citealp{Pra01}) mostly producing soft X-rays (\citealp{GudJ07}), or outflows launched in the region between the star and the disk as predicted in the X-wind model (\citealp{Shu00}).
 Non-accreting weak-line T Tauri stars (WTTS) are found to be more X-ray luminous than classical accreting T Tauri stars (CTTS) (\citealp{Ste01,Fla03a},2005), because CTTS absorb part of the X-rays in the accretion column \citep{Gre08}.
 
The effects of stellar X-rays on protoplanetary disks around T Tauri stars have been studied by many groups: \citet{Gla04,Gla07} found X-rays to be important in heating the surface layers of the disk and estimated the strength of the fine-structure emission lines of [Ne\,{\sc ii}] and [Ne\,{\sc iii}], which arise from the warm disk atmosphere exposed to X-ray radiation. \citet{Nom07} focused on $\mathrm{H_2}$ level populations and line emission, finding X-rays to control $\mathrm{H_2}$ level population pumping together with UV.  Following work of \citet{Mei08}, \citet{Erc08} and \citet{Gla09} studied atomic-line diagnostics of the inner regions of protoplanetary disks and formation of water in the warm disk atmosphere, respectively. \citet{Erc08} and subsequent work focused on the X-ray role in causing photo-evaporation of the surface layers of the disk. \citet{Gor04,Gor08} include X-rays to predict [Ar\,{\sc ii}, [Ne\,{\sc ii}], [Fe\,{\sc i}], [S\,{\sc i}], [Fe\,{\sc ii}] and [Si\,{\sc ii}] as good indicators of gas physics in the disk. \citet{Woo09} studied carbon isotope fractionation using the same method as \citet{Gor04}  to calculate X-ray ionization rates. Recent work of \citet{Hen10} studied the role of UV and X-rays on $\rm C_2H$ column densities and excitation conditions in disks around T Tauri stars. 

The scope of this paper is to take a step back and first perform a study of the impact of X-rays on the 2D disk structure, on molecular ionized species and on observational diagnostics such as [O\,{\sc i}], [C\,{\sc ii}] and [N\,{\sc ii}]. These molecular species were not studied in the above listed previous papers, but are of fundamental interest in the light of their recent detections with the Herschel satellite \citep{Ben10,Bru10}. We undertake an exploratory study in this letter of the relative effects of UV and X-rays on the protoplanetary disks hydrostatic, thermal and chemical structure using a series of X-ray models with different $L_{\rm X}$ (X-ray luminosity).
\begin{table}[t!]
\caption{Examples of molecular dissociation and cross section calculation: CO dissociation can end up in three different channels which are summed up to give the total CO cross section, CH is assumed to follow only one path.}
\begin{tabular}{l|l|l|l}
\hline
Molecule & Product 1 & Product 2 & Cross Section\\
\hline
\hline
CO & $\rm C^{2+}$ & O & $\sigma_{\rm CO} = 1/3\sigma_{\rm C}$ +\\
   & $\rm C^+$ & $\rm O^+$  & $1/3(0.5\sigma_{\rm C} + 0.5\sigma_{\rm O})$ +\\
   & $\rm C$ & $\rm O^{2+}$ & $1/3\sigma_{\rm O}$ \\
\hline
CH & $\rm C^{2+}$ & H & $\sigma_{\rm CH} = \sigma_{\rm C}$  \\
\hline
\end{tabular}
\begin{tabular}{l}
Molecules included: $\rm H_2$, CH, NH, OH, CN, CO, $\rm N_2$, SiH, \\
 NO, $\rm O_2$, SiO, $\rm CH_2$, $\rm NH_2$, $\rm H_2O$, HCN, $\rm CO_2$
\end{tabular}
\label{mol}
\end{table} 
 
\section{Model}
The chemo-physical disk modeling code ProDiMo \citep{Woi09,Kam10} has been updated with X-ray physics and chemistry.
\subsection{Input spectrum}
The incident stellar spectrum used in \citet{Woi09} is composed of a solar model with $T_{\rm eff} = 5\,800$ K and the cromospheric fluxes of HD 129\,333 \citep{Dor94}. The UV luminosity between 91.2 and 205 nm is $L_{\rm UV} \sim 4\cdot 10^{31}$ erg s$^{-1}$. Different models for the X-ray input spectrum are adopted in the literature: \citet{Nom07} use a two temperature thin thermal plasma model to fit the observed TW Hydrae spectrum; \citet{Erc08} built synthetic coronal spectra using line and continuum emissivities from the CHIANTI compilation of atomic data; \citet{Gor08} use a thermal blackbody. We place the X-ray source on the star and follow \citet{Gla97}, \citet{Ige99} in using an analytic input spectrum to describe the Bremsstrahlung emission from an isothermal plasma: $F(E) \propto 1/E\cdot \mathrm{exp}(-E/kT_{\rm X})$ where E is the photon energy between 0.1 and 100 keV, $kT_{\rm X} = 1\,\rm{keV}$, and $T_{\rm X}$ is the plasma temperature. The X-ray input spectrum is added to the input spectrum shown in Fig. 2 in \citet{Woi09}. 
\subsection{X-Ray chemistry}
The following paragraphs describe the implementation of primary and secondary ionization in the code. 41 primary ionization reactions and 16 secondary ionization reactions were added in the chemical network.

\subsubsection*{Primary Ionization}
The primary rates for X-ray absorption are calculated following \citet{Mei05} (Appendix D.3.1). The cross sections are taken from \citet{Ver95}. Since X-rays are likely absorbed in the K-shells, we assume that every X-ray absorption leads to a single ionization for H, He, Si, Cl and a double ionization for C, N, O, S and Fe \citep{Mei05}. We take into account molecular X-ray absorption, which leads always to dissociation, of the species. Table \ref{mol} lists two examples of how the total dissociation rates are calculated. When the difference in weight of the elements that form the molecule is large  (e.g. CH, OH, $\rm H_2O$ etc.), we use the cross section of the heavier element, when the weight is comparable (e.g. CO, CN, NO etc.) we combine the cross sections of the single elements.

\subsubsection*{Secondary Ionization}
The fast electrons generated by the X-ray absorption can further ionize the gas. The rates are computed as shown in Meijerink \& Spaans (2005, Appendix D.3.2) using experimental data from \citet{Len88}. The volumetric rates are a function of the local chemistry, i.e. they depend on the species densities $n_{\rm H}$, $n_{\rm H_2}$, $n_{\rm <H>}$ and $n_{\rm e}$ (where $n_H$ is the hydrogen atom density and $n_{\rm <H>}$ is the hydrogen nuclei density). This has to be taken into account when solving the chemical equilibrium.  The additional entries in the chemical Jacobian that come from  the X-ray secondary ionization reactions are implemented following Sect. 5.6 in \citet{Woi09}.
\subsection{X-Ray heating}
We added Coulomb heating \citep{Dal99} and $\rm H_2$ ionization heating (Meijerink \& Spaans, 2005, Appendix B.1) to the heating processes listed in \citet{Woi09}. The \citet{Mal96} heating rate only holds for gas with low $x_e$, while \citet{Dal99} present the more general case for different high and low ionization gases (atomic/molecular). The heating efficiency increases by a factor 7-8 in a highly ionized atomic gas and a factor 2 in a highly ionized molecular gas \citep{Shu85}.
\subsection{Parameter space}
 We computed five different models with increasing $L_{\rm X}$ to compare the results with the UV only case ($\mathrm{\mathtt{Model\,1}}$) for a disk surrounding a T Tauri star (Table \ref{mod}).
\begin{small}
   \begin{table}[t]
\centering
\caption{Parameters used in the models: the X-ray luminosity values correspond to $\mathtt{Models}$ from $\mathtt{1}$ to $\mathtt{5}$.}
\begin{tabular}{l|c|c}
\hline
Quantity & Symbol & Value \\
\hline
\hline
Stellar mass       & $M_*$      & 1 $M_{\odot}$ \\[1mm]
Stellar luminosity & $L_*$      & 1 $L_{\odot}$ \\[1mm]
Disk mass          & $M_{\rm disk}$ & 0.01 $M_{\odot}$ \\[1mm]
X-ray luminosity (0.1-50 keV)& $L_{\rm X}$   & $0,10^{29},10^{30}$\\ 
 &    & $10^{31},10^{32}$ \\[1mm]
Inner disk radius  & $R_{\rm in}$   & 0.5 AU \\[1mm]
Outer disk radius  & $R_{\rm out}$   & 500 AU \\[1mm]
Surface density power law index& $\epsilon$ & 1.5 \\[1mm]
Min. dust particle size & $a_{\rm min}$ & 0.1 $\rm{ \mu m}$ \\[1mm]
Max. dust particle size & $a_{\rm max}$ & 10 $\rm{ \mu m}$ \\[1mm]
Dust size distribution power index& $a_{\rm pow}$ & 2.5 \\[1mm]
\hline
\end{tabular}
\label{mod}
\end{table}
 \end{small}
\noindent In {$\mathtt{Model\,1}$} X-rays are switched off. This is the model as presented in \citet{Woi09}. For more details, we refer to the original paper. X-ray models ($\mathtt{Model\,2-5}$) have the same parameters as $\mathtt{Model\, 1}$, but include stellar X--rays ($L_{\rm X} = 10^{29}-10^{32}\,\mathrm{erg\,s^{-1}}$). We added Cl, $\mathrm{Cl^+}$ and double ionized species ($\mathrm{C^{2+}}$, $\mathrm{N^{2+}}$, $\mathrm{O^{2+}}$, $\mathrm{S^{2+}}$, $\mathrm{Fe^{2+}}$) to the chemical network. The X-ray emission is treated as a point source at the location of the star.

\begin{figure*}[t!]
  \centering
  \includegraphics[scale=0.35]{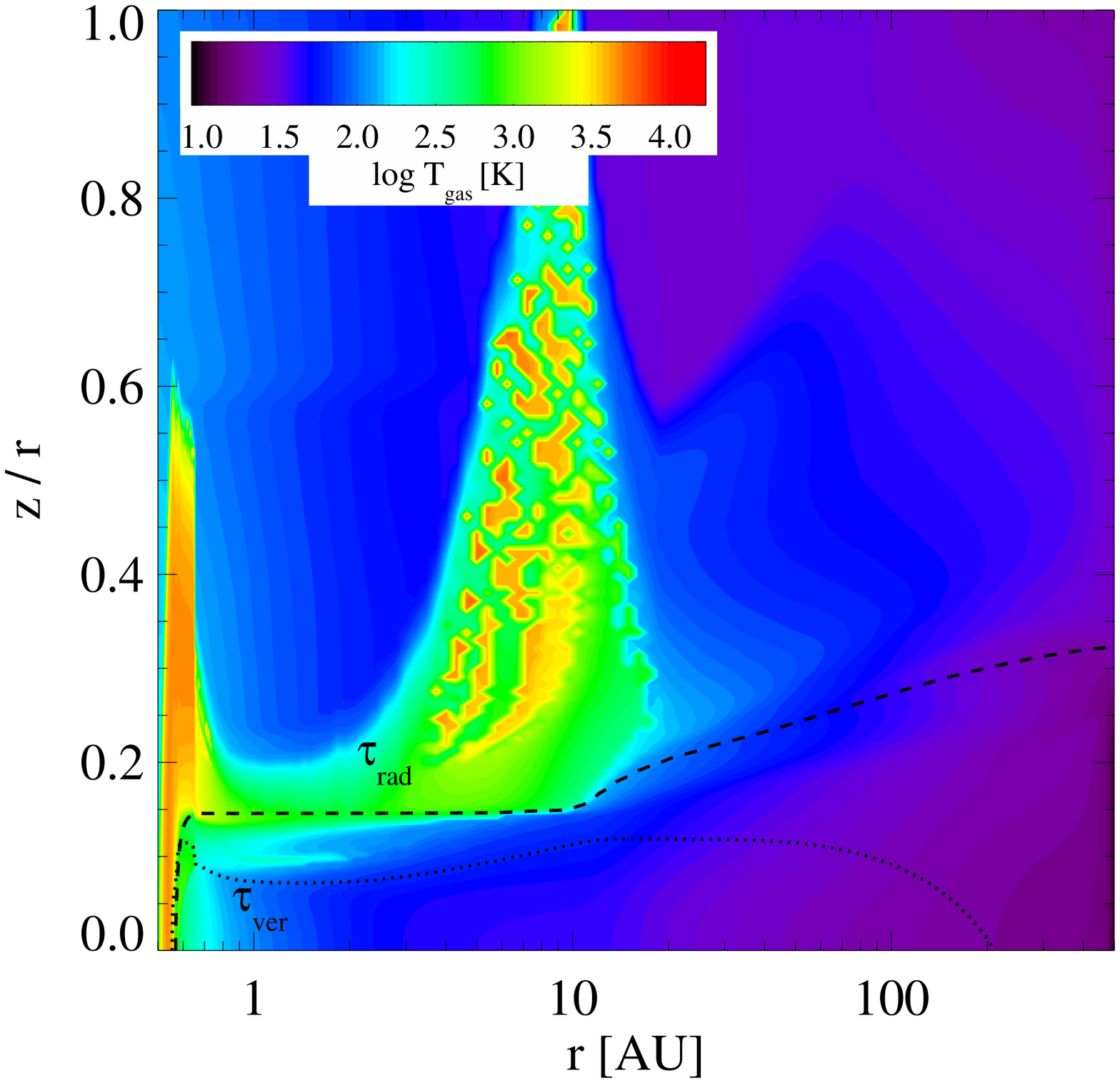}%
  \includegraphics[scale=0.35]{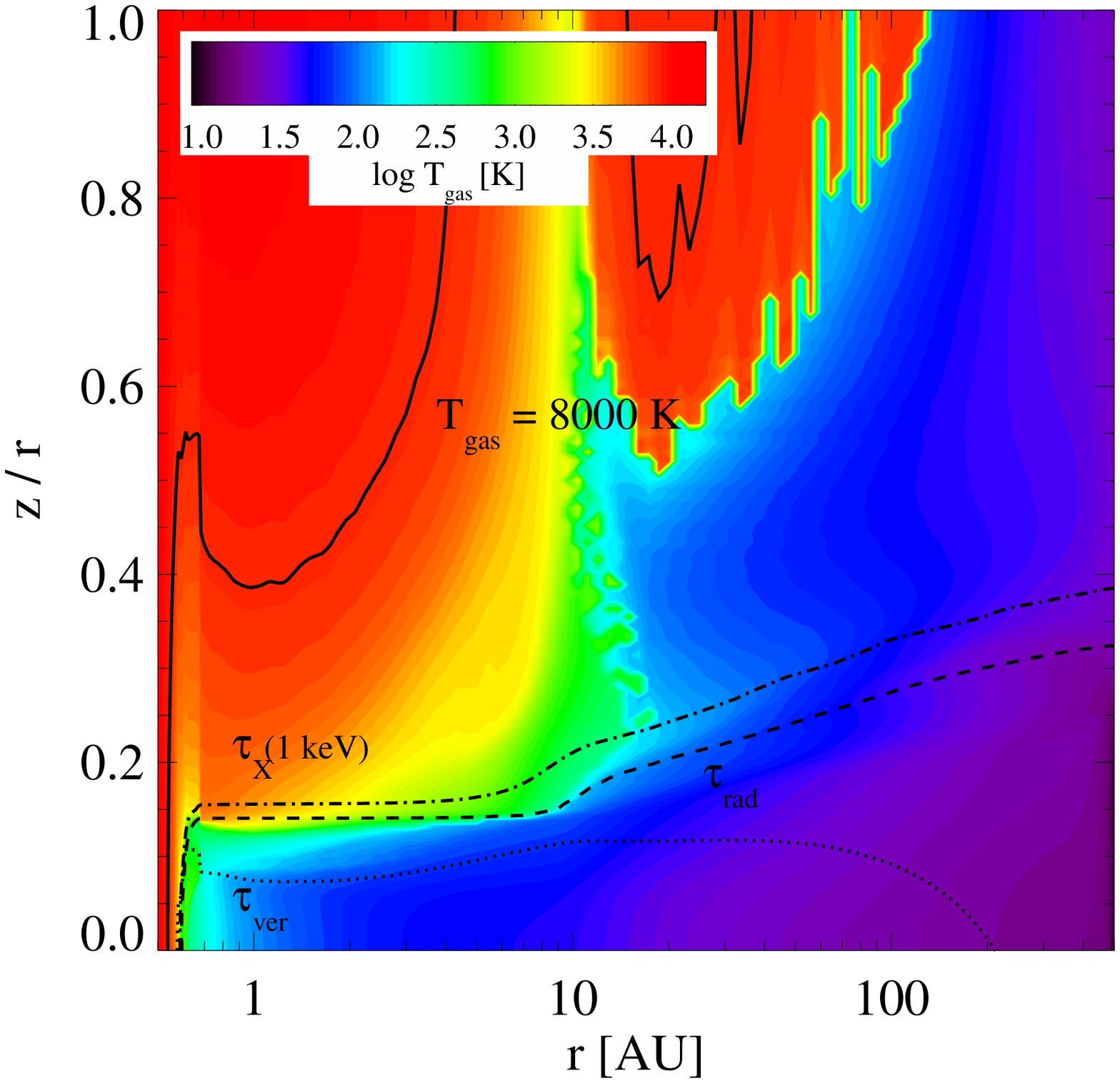}%
  \includegraphics[scale=0.35]{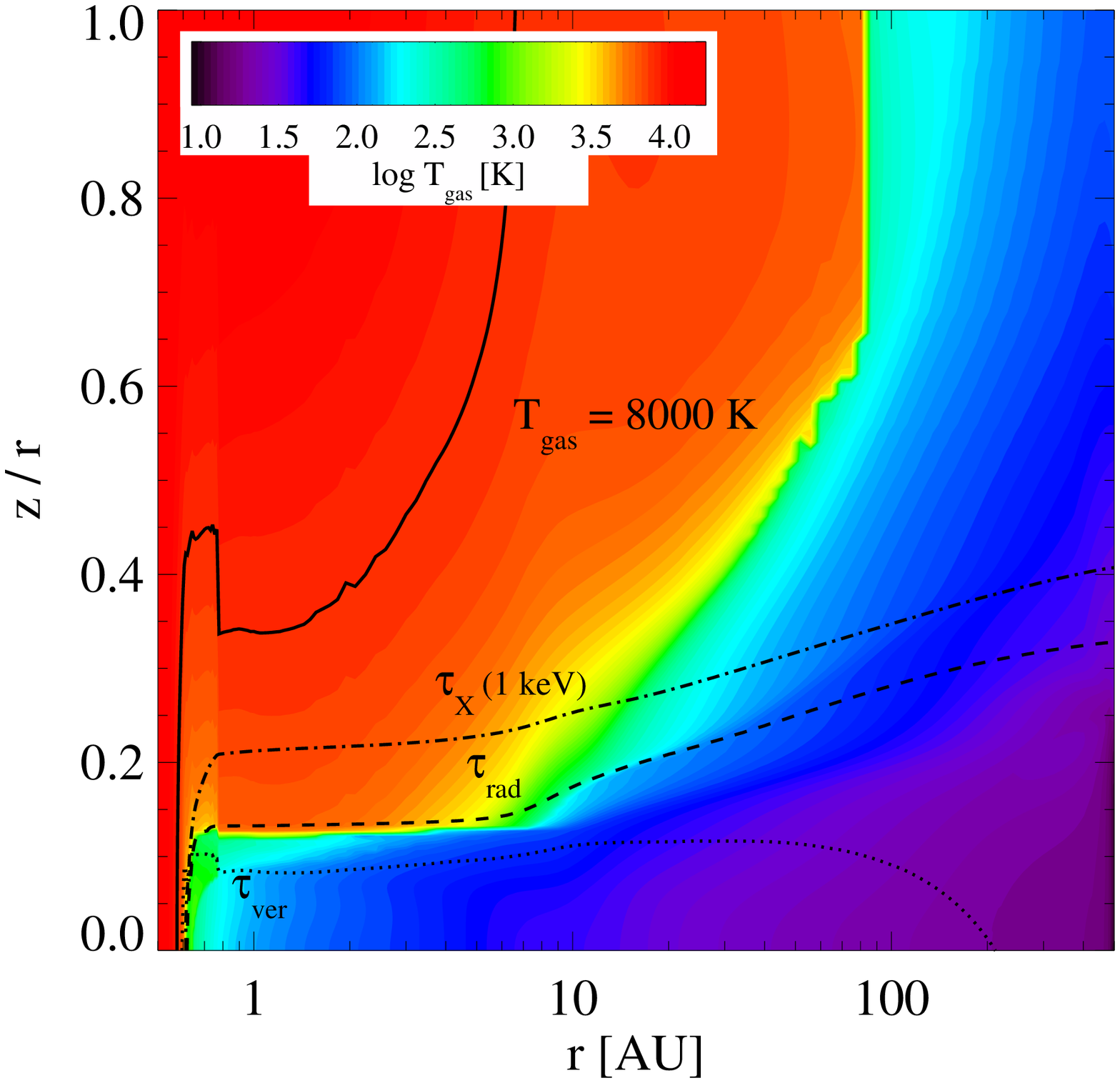}

  \caption{Gas temperature distribution: $\mathtt{Model\,1}$ (UV only) in the left panel, X-ray models with $L_{\rm X}=10^{30}\mathrm{erg\,s^{-1}}$ ($\mathtt{Model\,3}$), $L_{\rm X}=10^{32}\mathrm{erg\,s^{-1}}$ ($\mathtt{Model\,5}$) in the middle and right panel respectively. Contour lines are over-plotted for $\mathrm{\tau_{rad} = 1}$ (radial dust optical depth at 550 nm, dashed line), $\mathrm{\tau_{ver} = 1}$ (vertical dust optical depth at 550 nm, dotted line) and X-ray optical depth of one at 1 keV (dot-dashed line). The relative position of these two depth depend strongly on the assumed dust properties (see Table \ref{mod}). The solid line corresponds to $T_{\rm gas} = 8\,000$ K.}
  \label{tpl}
  \end{figure*}
\begin{figure*}[t!]
  \centering
  \includegraphics[scale=0.35]{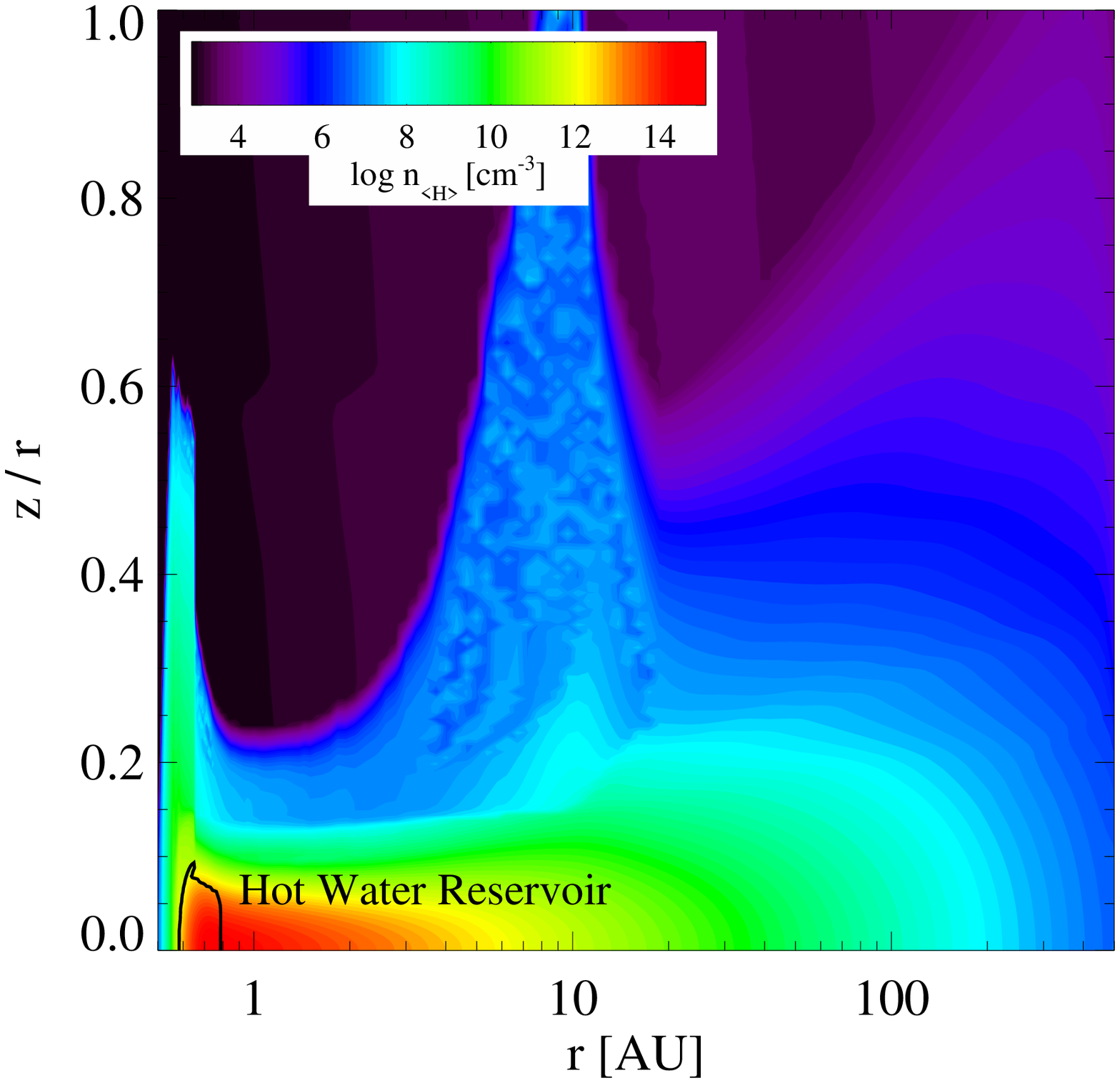}%
  \includegraphics[scale=0.35]{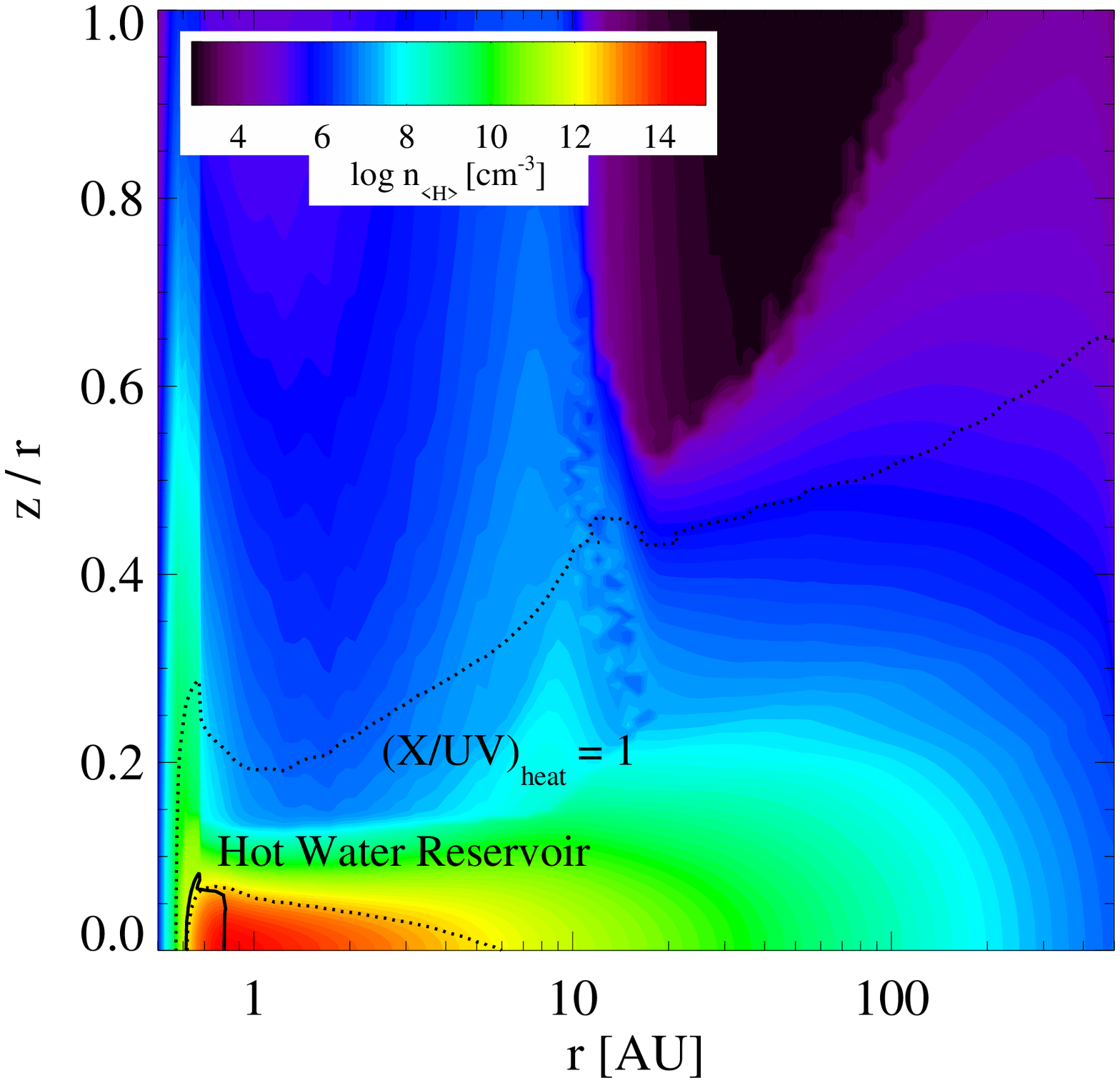}%
  \includegraphics[scale=0.35]{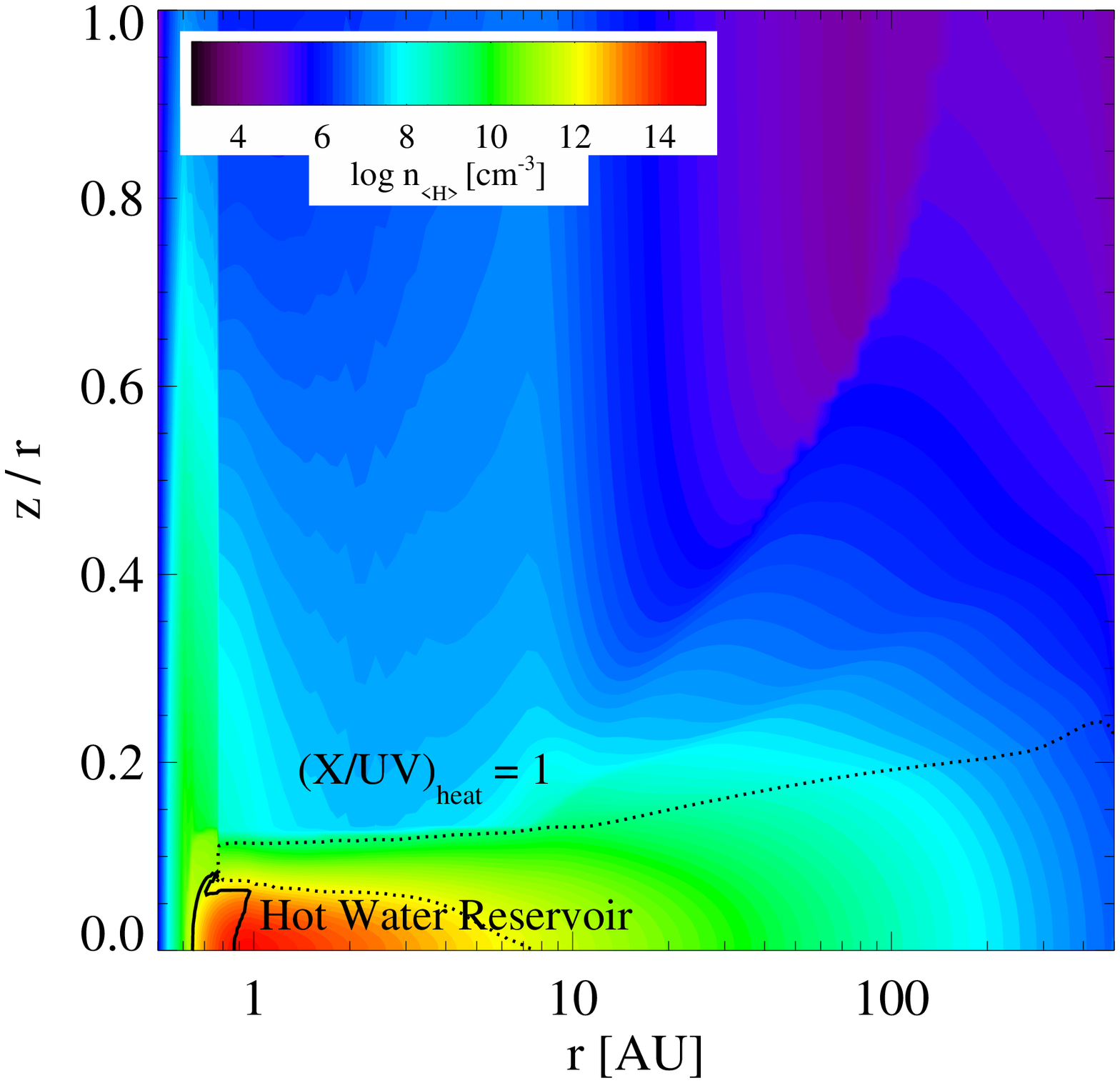}
  \caption{Total hydrogen nuclei density $n_{\rm <H>}$ distribution: $\mathtt{Model\,1}$ (UV only) in the left panel, X-ray models with $L_{\rm X}=10^{30}\mathrm{erg\,s^{-1}}$ ($\mathtt{Model\,3}$), $L_{\rm X}=10^{32}\mathrm{erg\,s^{-1}}$ ($\mathtt{Model\,5}$) in the middle and right panel respectively. The dotted contour line shows where the ratio between Coulomb heating and photo-electric plus PAH heating equals 1. The solid line circles the hot water reservoir ($T_{\rm H_2O} >$ 200 K).}
 \label{dpl}
\end{figure*}
\section{Results}

We analyze the results obtained from the five models described above. We first present the thermal and hydrostatic structure of the various models. Then we describe the impact of X-rays on the chemistry and fine structure line emission, focusing on the main differences between the UV only model and the combined UV+X-ray models.
\subsubsection*{Temperature and density structure}
The low density ($n_{\rm <H>} < 10^8\,\rm cm^{-3}$) surface layers directly exposed to X-ray radiation are heated efficiently via Coulomb heating: in all X-ray models temperatures reach up to $\sim$8\,000 K (Fig. \ref{tpl}). The extension of this high temperature region increases with $L_{\rm X}$: it reaches 2 AU in $\mathtt{Model\,2}$ and 100 AU in $\mathtt{Model\,3}$ and $\mathtt{Model\,4}$. In $\mathtt{Model\,5}$ the temperature is slightly lower for $r >$ 50 AU but the whole disk is warmer (Fig. \ref{tpl}). X-rays dominate the thermal balance only in the upper layers down to $\mathrm{(X/UV)_{heat} = 1}$. Beyond $\mathrm{(X/UV)_{heat} = 1}$, UV heating takes over and becomes the main heating process. Deeper in the midplane, both X-ray and UV heating are negligible, but their ratio is inverted again and harder X-rays ($E > 10$ keV) deposit more energy than UV radiation (Fig. \ref{dpl}). 

The inner wall of the disk is always directly illuminated by the stellar radiation causing the inner rim (0.5 AU $< r <$ 0.7 AU) to have a large scale height and hence vertical extent. In the UV model the shadowed disk atmosphere beyond the inner rim has a lower vertical scale height since radiation cannot penetrate efficiently and sustain high gas temperature. Only radiation that impinges the disk at higher angles ($z/r > 0.6$) and scattered radiation reach the outer disk causing the temperature to increase and consequently the second bump at $\sim$ 9 AU. When X-rays are switched on, the temperature increase in the surface layers affects directly the vertical density distribution there since we keep the radial surface density profile fixed. The inner rim of the disk ($r < 1 $ AU and $z/r > 0.1$) is more puffed up (Fig. \ref{dpl}). The second bump at $\sim$ 9 AU is more vertically extended as well; in $\mathtt{Model\,5}$ it even ''merges'' with the inner rim and flares as a whole (Fig. \ref{dpl}, third panel). 
\subsubsection*{Chemistry}
\begin{figure}[t]
  \centering
  \includegraphics[angle=-90,trim=-20 0 0 0,scale=0.36]{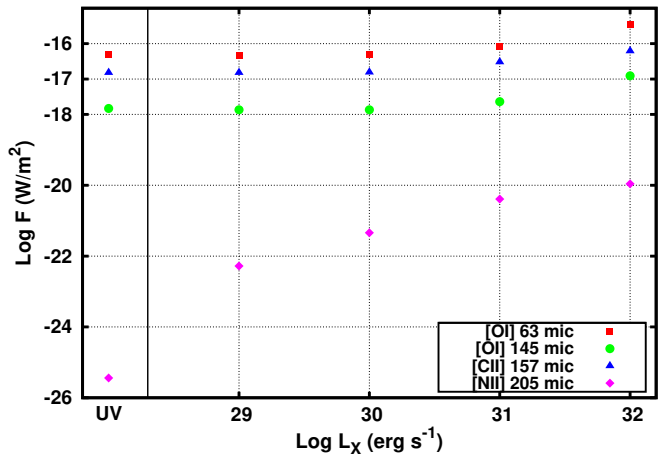}
  \caption{Oxygen line flux emission at 63 and 157 $\rm \mu m$, ionized carbon line flux emission at 157 $\rm \mu m$ and ionized nitrogen line flux emission at 205 $\rm \mu m$ as a function of the X-ray luminosity. All fluxes are compiled for a distance of 140 pc. The correlation between N\,{\sc ii} and $L_{\rm X}$ is clear, while O\,{\sc i} and C\,{\sc ii} show no correlation for $L_{\rm X} <$ 10$^{30}$ erg s$^{-1}$.}
  \label{lines}
\end{figure}
\begin{figure}[t]
  \centering
  \includegraphics[trim=0 30 -40 0,scale=0.7,width=7cm,height=9cm,angle=-90]{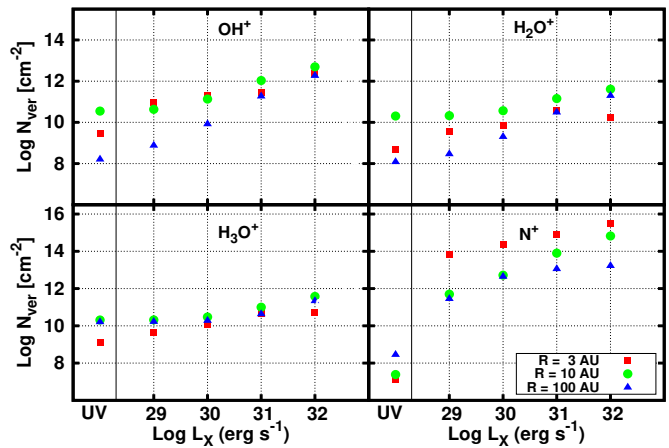}
  
  \vspace{0.3cm}
  \caption{Column density of $\mathrm{OH^+}$ (upper left panel), $\mathrm{H_2O^+}$ (upper right panel), $\mathrm{H_3O^+}$ (lower left panel) and $\mathrm{N^+}$ (lower right panel) at three different radii (3, 10, 100 AU) versus X-ray luminosity. In each panel the UV only model is indicated with the ''UV'' label on the x-axis.}
  \label{nvsl}
\end{figure}

The abundances of neutral molecules such as $\mathrm{CO}$ and $\mathrm{OH}$ and $\rm H_2O$ generally change by less than an order of magnitude over the entire modeling space. On the other hand, the ion chemistry in the disk surface is strongly changed: the electron abundance in the upper layers increases  by a factor $10^4$. This leads to an enhancement of $\mathrm{OH^+}$, $\mathrm{H_2O^+}$, $\mathrm{H_3O^+}$ and $\mathrm{N^+}$ abundances. Secondary ionization of $\rm H_2$ increases the $\rm H_2^+$ abundances. $\rm H_2^+$ collisions with $\rm H_2$ produce $\rm H_3^+$. Subsequent collisions of $\rm H_3^+$ with O, OH and $\rm H_2O$ lead to the production respectively of $\mathrm{OH^+}$, $\mathrm{H_2O^+}$ and $\mathrm{H_3O^+}$. Fig. \ref{nvsl} shows how the vertical column densities of these species change with $L_{\rm X}$ for three different disk radii: 3, 10 and 100 AU. $\rm OH^+$ and $\rm H_2O^+$ respond more efficiently than $\rm H_3O^+$ to the X-ray radiation as the total mass ratio $\rm M(OH^+)$/$\rm M(H_2O^+)$ is only doubled from $\mathtt{Model\,1}$ to $\mathtt{Model\,5}$, while $\rm M(OH^+)$/$\rm M(H_3O^+)$ increases by more than a factor 5.
\\
\linebreak
$\mathbf{OH^+}$:
$\mathrm{OH^+}$ formation is enhanced approximately proportional to the X-ray luminosity. Its total mass increases from $8 \times 10^{-15}\, M_{\odot}$ in $\mathtt{Model\,1}$ by a factor 3.6 in $\mathtt{Model\,3}$ and about 500 times in $\mathtt{Model\,5}$ (Table \ref{mt}). 


As the X-ray luminosity increases, $\mathrm{OH^+}$ formation is enhanced and pushed outwards, where the disk is colder. The column density of $\mathrm{OH^+}$ increases substantially beyond $r > 10$ AU in all models. It reaches $\sim 100$ times its initial value ($\mathtt{Model\,1}$) in $\mathtt{Model\,3}$ at 100 AU (Fig. \ref{nvsl}).
\\
\linebreak
$\mathbf{H_2O^+}$:
The $\mathrm{H_2O^+}$ mass increases by 3 times (factor of 100) between $\mathtt{Model\,1}$ and $\mathtt{Model\,3}$ ($\mathtt{Model\,5}$). The column density again increases toward the outer part of the disk. $\mathtt{Model\,3}$ has $\sim10$ times higher column density than $\mathtt{Model\,1}$ at 10 and 100 AU. As in the case of  $\mathrm{OH^+}$, $\mathrm{H_2O^+}$ is pushed toward the outer part of the disk with increasing X-ray luminosity.
\\
\linebreak
$\mathbf{H_3O^+}$: The $\mathrm{H_3O^+}$ mass is slightly less affected by X-rays: it is 2 times higher than in the UV case in $\mathtt{model\,3}$ and 10 times in $\mathtt{model\,4}$ . In $\mathtt{model\,3}$, $\mathtt{model\, 4}$ and $\mathtt{model\, 5}$, $\mathrm{H_3O^+}$ is also clearly pushed to larger radii and lower temperatures. 
\\
\linebreak
$\mathbf{N^+}$:
We find the most extreme mass increase for $\mathrm{N^+}$ in all X-ray models (Table \ref{mt}) and at all radii: in $\mathtt{model\,3}$ it is 1\,000 times higher than in the UV only case. 

\begin{figure}[htbp]
  \centering
  \includegraphics[angle=-90,trim=-20 0 0 0,scale=0.35]{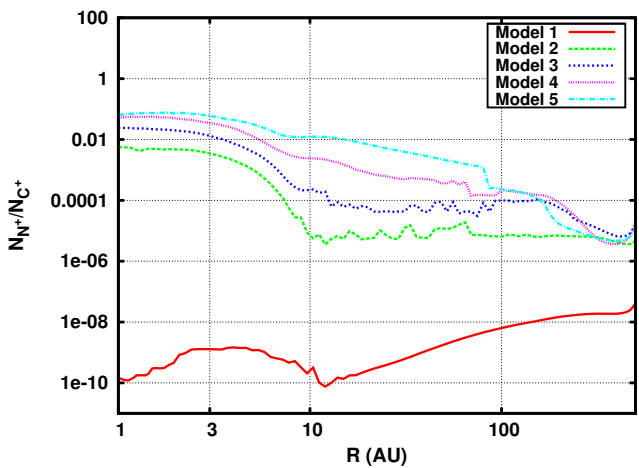}
  \caption{Column density ratio between the X-ray dominated $\mathrm{N^+}$ and the UV dominated $\mathrm{C^+}$. The latter does not change in the X-ray models with respect to the UV model}
  \label{NC}
\end{figure}

In $\mathtt{models\, 4}$ and $\mathtt{5}$, the $\rm N^+$ mass increases over-proportionally in the outer part of the disk (Fig. \ref{nvsl}, right lower panel). Figure \ref{NC} shows the column density ratio of the X-ray dominated $\rm N^+$ to the UV dominated $\rm C^+$. With increasing $L_{\rm X}$, the $\rm N^+$ column density is enhanced by two (outer disk) to six (inner disk) orders of magnitude.
\subsection*{Line Predictions}
In table \ref{lines} we list the flux prediction of [O\,{\sc i}] 63 $\rm{ \mu m}$, [O\,{\sc i}] 145 $\rm{ \mu m}$, [C\,{\sc ii}] 157 $\rm{ \mu m}$ and [N\,{\sc ii}] 205 $\rm{ \mu m}$ for all the models. Fig. \ref{lines} shows their values as a function of the X-ray luminosity. The [N\,{\sc ii}] line flux shows the strongest correlation with the X-ray flux, increasing by 5 orders of magnitude from $\mathtt{model\,1}$ to $\mathtt{model\,3}$. The oxygen fine structure line fluxes and the [C\,{\sc ii}] line flux are constant from $\mathtt{model\,1}$ to $\mathtt{3}$ (fig. \ref{lines}), due to the fact that they originate in the UV heated layer. Only in $\mathtt{model\,4}$ and $\mathtt{5}$ the line fluxes increase with $L_{\rm X}$.
\begin{center} 
\begin{table*}[ht!]
\centering
\caption{Predicted line fluxes of [O\,{\sc i}] at 63 $\rm{ \mu m}$, [C\,{\sc ii}] at 157 $\rm{ \mu m}$ and [N\,{\sc ii}] at 205 $\rm{ \mu m}$ for all the models expressed in $\rm W/m^2$. All fluxes are computed for a distance of 140 pc.}
\begin{tabular}{l|c|c|c|c|c}
\hline
Line & $\mathtt{Model\, 1}$ & $\mathtt{Model\,2}$ & $\mathtt{Model\, 3}$ & $\mathtt{Model\, 4}$ & $\mathtt{Model\, 5}$\\
\hline
\hline
$[$O\,{\sc i}$]$ 63 $\rm{ \mu m}$    & 4.8(-17) & 4.6(-17) & 4.8(-17) & 8.3(-17) & 3.6(-16)  \\
$[$O\,{\sc i}$]$ 145 $\rm{ \mu m}$    & 1.5(-18) & 1.3(-18) & 1.4(-18) & 2.3(-18) & 1.2(-17)  \\
$[$C\,{\sc ii}$]$ 157 $\rm{ \mu m}$ & 1.5(-17) & 1.5(-17) & 1.7(-17) & 3.0(-17) & 6.1(-17) \\
$[$N\,{\sc ii}$]$ 205 $\rm{ \mu m}$ & 3.6(-26) & 5.2(-23) & 4.6(-22) & 4.1(-21) & 1.1(-20) \\
\hline
\end{tabular}
\label{lines}
\end{table*}
\end{center}

\begin{table*}[ht!]
\centering
\caption{Species masses given in units of solar mass in all the models}
\begin{tabular}{l|c|c|c|c|c}
\hline
Species & UV Only & $L=10^{29}\mathrm{erg\,s^{-1}}$  & $L=10^{30}\mathrm{erg\,s^{-1}}$  & $L=10^{31}\mathrm{erg\,s^{-1}}$ &$L=10^{32}\mathrm{erg\,s^{-1}}$ \\
\hline
\hline
$\mathrm{H}$     & $5.0(-05)$  & $5.0(-05) $  &   $5.2(-05)$    &   $7.1(-05) $ & $1.4(-04)$  \\
$\rm H_2$        & $7.6(-03) $ & $7.6(-03) $  &   $7.6(-03)   $ &   $7.6(-03)$ & $7.5(-03) $\\
$\mathrm{C}$     & $3.0(-07)$  & $3.0(-07) $  &   $3.1(-07)$    &    $3.4(-07) $ & $4.8(-07)$  \\
$\mathrm{C+}$    & $1.3(-07)$  & $1.3(-07) $  &   $1.3(-07)$    &   $1.4(-07) $ & $1.8(-07)$  \\
$\mathrm{O}$     & $1.4(-05)$  & $1.4(-05) $  &   $1.4(-05)$    &   $1.4(-05) $ & $1.4(-05)$  \\
$\mathrm{CO}$    & $9.5(-06)$  & $9.4(-06) $  &   $9.3(-06)$    &   $9.0(-06) $ & $ 9.0(-06)$ \\
$\mathrm{OH}$    & $3.4(-11)$  & $3.4(-11) $  &   $3.7(-11)$    &   $7.0(-11) $ & $2.8(-10)$  \\
$\mathrm{H_2O}$  & $7.7(-07)$  & $4.7(-08) $  &   $3.6(-08)$    &   $2.7(-08) $ & $3.3(-08)$  \\
$\mathrm{OH^+}$  & $7.9(-15) $ & $1.1(-14) $  &   $2.9(-14)$    &   $4.1(-13) $ & $4.6(-12) $ \\
$\mathrm{H_2O^+}$& $1.9(-15)$  & $2.1(-15) $  &   $5.3(-15)$    &   $6.2(-14) $ & $4.2(-13)$  \\
$\mathrm{H_3O^+}$& $3.4(-15)$  & $3.7(-15) $  &   $7.3(-15)$    &   $4.8(-14) $ & $3.5(-13)$  \\
$\mathrm{N^+}$   & $2.9(-15)$  & $8.7(-13) $  &   $3.9(-12)$    &   $9.9(-12) $ & $ 3.6(-11)$ \\
\hline
\end{tabular}
\label{mt}
\end{table*}
\section{Discussions and conclusions}
We find that X-rays mainly impact the surface layers of protoplanetary disks. The hydrostatic structure of these layers is significantly changed: X-ray radiation is mostly absorbed in the tenuous layers ($z/r \geq 0.2\,\rm cm^{-3}$) increasing the temperature up to $\sim 8\,000$ K; at those temperatures cooling by [O {\sc I}], Fe {\sc II} and Ly$\alpha$ balance the X-ray heating. This causes the density structure to flare stronger than in the UV only case (Fig. \ref{dpl}).

The general results from \citet{Nom07} and \citet{Gla04} are qualitatively confirmed, showing that the inner rim and the surface layers are dominated by X-rays. In addition we find that more distant parts, up to 100 AU, are affected as well (Fig. \ref{nvsl}). We find that the size of the high temperature X-ray driven region increases with $L_{\rm X}$. This is due both to the fact that our model uses the \citet{Dal99} prescription for the X-ray heating and to our choice of the input spectrum. The \citet{Dal99} heating efficiencies lead to $\sim 7-8$ times higher rates in regions with high electron fraction, compared to \citet{Mal96}. Furthermore our input spectrum has higher photon flux in the range 0.1-0.3 keV compared with, e.g., \citet{Gor08} and \citet{Nom07}. These soft X-ray photons are absorbed at low vertical column densities causing high $\rm e^-$ densities there and thus are more efficient in heating the upper layers. This effect was also noted in \citet{GDH09}.

\citet{Gor08} found X-rays to be the dominant heating process in slightly deeper layers than we do. This is most likely due to the fact that we include UV scattering from dust, which leads to a larger vertical penetration of FUV radiation in the disk. \citet{Nom07} also show that FUV penetrates deeper than X-rays, because FUV scattering is more efficient than X-ray Compton scattering at 1 keV, which is not included in our model. \citet{Gor08} also consider a less steep surface density distribution which yields less mass at 1 AU, where most of the X-ray radiation is absorbed in our model.

The X-ray energy deposition does not affect neutral oxygen fine structure emission at 63 and 145 $\rm \mu m$ (Table \ref{lines}), unless the X-ray luminosity is pushed to $L_{\rm X} > 10^{30}\,\rm erg\,s^{-1}$, where X-ray heating dominates in the regions where the emission takes place. Ionized carbon emission at 157 $\rm{ \mu m}$ is only slightly affected by the X-rays in all the models, as the line emission is dominated by material at R $>$ 200 AU \citep{Woi09,Kam10}, where the X-ray contribution to the chemistry and thermal balance is negligible. Our values are in good agreement with \citet{Mei08} and \citet{Erc08} for $L_{\rm X} \sim 10^{30}$ erg s$^{-1}$. However the correlation of the oxygen line fluxes with $L_{\rm X}$, found in the \citet{Mei08} model is due to the lack of UV radiation in their model. In our model at low $L_{\rm X}$, the UV heating provides the gas temperature to sustain a constant level of [O\,{\sc i}] emission.

We find $\mathrm{H_2}$, CO and OH properties in the X-ray models largely unvaried with respect to $\mathtt{Model\,1}$. The largest change occurs for water which mass diminishes by a factor 20 in the highest $L_{\rm X}$ model compared with the UV only model. 
The X-rays enhance the electron abundance in the surface layers, where we find $x_{e} \sim 0.1$ for $z/r \sim 0.9$ out to 100 AU. Ionized species abundances are enhanced, such as $\mathrm{OH^+}$, $\mathrm{H_2O^+}$, $\mathrm{H_3O^+}$ and $\mathrm{N^+}$. The molecular ionized species formation is pushed outwards as $L_{\rm X}$ increases. High abundances of  $\mathrm{OH^+}$ and $\mathrm{H_2O^+}$ are a characteristic signature of an X-ray dominated region as recently discussed by \citet{vdw10} after the detection with Herschel/SPIRE in the ultra-luminous galaxy Mrk 231. Furthermore this is extremely interesting in the context of the recent detection of these species in a massive YSO by Herschel \citep{Ben10}. The column densities that we find from our X-ray irradiated disks are of the same order of magnitude as those derived in \citet{Ben10}. This suggests that a disk could in principle bring a substantial contribution to the line fluxes emitted from these molecules, making our result extremely relevant for future Herschel studies of low mass YSOs.

The abundance of these ions increases especially at larger radii ($r > 10$ AU), while the $\mathrm{N^+}$ abundance rises considerably at all radii. The latter is not observed in the UV model due to the high ionization potential of $\rm N(I.P.) =$ 14.5 eV. We find that the $\rm{N^+}$ flux emitted at 205 $\rm{ \mu m}$ in $\mathtt{model\,3}$ is $\sim 5\times 10^{-22}\,\rm{W/m^2}$, coming for 90\% from the upper layers beyond 10 AU. This is much smaller than the current Herschel sensitivity limit of $5\times 10^{-18}\,\rm W/m^2$ (HIFI, $5\,\sigma$ in 1 hour) and even below current sensitivity estimates for SPICA/SAFARI ($10^{-19}$ W/m$^2$, $5\,\sigma$ in 1 hour). We only consider X-ray ionization, but the flux is unlikely to change if EUV is present. Additional ionization occurs only at smaller radii, because of the low penetration depth of EUV radiation \citep{Hol09}. The molecular ionized species formation is pushed toward the outer disk as $L_{\rm X}$ increases.

\section{Outlook}
The predicted $\mathrm{N^+}$ line fluxes are too low to be observed with current instruments. In contrast, the [O\,{\sc i}] 63 $\rm \mu m$ and [C\,{\sc ii}] 157 $\rm \mu m$ lines are observable in the presence of UV luminosities of $\sim$ 10$^{31}$ erg s$^{-1}$ and/or strong X-rays ($L_{\rm X} > 10^{30}$ erg s$^{-1}$). In the context of the current Herschel observatory, the $\rm OH^+$, $\rm H_2O^+$ and $\rm H_3O^+$ lines are extremely interesting as they are likely observable with the HIFI, PACS and SPIRE instruments. These hydride ions and the already observed [Ne {\sc II}] and [Ar {\sc II}] line fluxes, will be addressed in the subsequent paper (Aresu et al., in preparation).  We will also investigate a larger parameter space and the effect of the change in disk structure - especially the higher inner rim and increased flaring - on the overall SED. Future work also includes the creation of a grid of X-ray models following the \citet{Woitke10} approach.

\noindent
\newline
\tiny \emph{Acknowledgments}. The authors acknowledge the anonymous referee for helpful comments that improved the clarity and completeness of the paper. This work has been funded by the Netherlands Organization for Scientific Research (NWO). The LAOG group acknowledge PNPS, CNES and ANR (contract ANR-07-BLAN-0221) for financial support.


\bibliographystyle{aa}
\bibliography{biblio}

\end{document}